\documentclass[twocolumn,letterpaper,aps,prc,superscriptaddress,nofootinbib,showpacs,floatfix]{revtex4-1}
\usepackage{comment}
\usepackage{setspace}
\usepackage{color}
\usepackage{indentfirst}
\usepackage{xspace}
\usepackage{verbatim}
\usepackage{epstopdf}
\usepackage[export]{adjustbox}
\usepackage[T1]{fontenc}
\usepackage{float}
\usepackage{wrapfig}
\usepackage{graphicx}
\usepackage{subcaption}
\usepackage{diffcoeff}
\usepackage{amsmath}
\usepackage{amssymb}
\usepackage{amsthm}
\usepackage{array}
\usepackage{tabularx}
\usepackage[utf8]{inputenc}
\graphicspath{ {figures/} }
\usepackage{lineno}
{\large }

\usepackage{natbib}
\setcitestyle{square, comma, numbers,sort \& compress, super}
\usepackage{appendix}
\usepackage{dcolumn}
\usepackage{bm}
\usepackage[colorlinks=true, linktocpage=true, linkcolor=blue, citecolor=blue]{hyperref}
\usepackage[a4paper]{geometry}
\begin{document}
\title{Charged particle multiplicity fluctuation in $A-A$ collisions at RHIC and LHC energies using  Angantyr model.}

\author{Pritindra Bhowmick}
\email{pritindra18@iiserb.ac.in}
\affiliation{Department of Physics, Indian Institute of Science Education and Research, Bhopal, India-462066 }

\author{ Sadhana~Dash }
\email{sadhana@phy.iitb.ac.in}
\affiliation{Department of Physics. Indian Institute of Technology Bombay, Mumbai, India-400076}

\author{ Basanta Kumar ~Nandi }
\email{basanta@phy.iitb.ac.in}
\affiliation{Department of Physics. Indian Institute of Technology Bombay, Mumbai, India-400076}

\author{ Claude ~Pruneau }
\email{pruneau@physics.wayne.edu}
\affiliation{Department of Physics and Astronomy, Wayne State University, Detroit, 48201, USA}

\begin{abstract}

Event-by-event fluctuations of the charged particle multiplicity  are studied for a wide range of centralities for Au$-$Au collisions at $\sqrt{s_{NN}}$ = 200 GeV,   
 Pb$-$Pb collisions at $\sqrt{s_{NN}}$ = 2.76 TeV and 5.02 TeV using the Pythia 8 Angantyr model.  The centrality dependence of $\omega_{ch}$ observable, which quantifies the fluctuations in terms 
of scaled variance  is studied for different pseudorapidity ranges and has been compared with those obtained from a simple  participant superposition model.  
The $\omega_{ch}$ was found to be lower than the expectations from the participant model. The estimate would act like a baseline for current and future measurements of 
 event-by-event fluctuations in the charged particle multiplicities in 
systems at LHC energies where no de-confined  medium of quarks and gluons are formed.
\end{abstract}

\maketitle

\section{Introduction}

Comparisons of measurements of anisotropic flow and transverse momentum correlator carried out in heavy-ion collisions at RHIC and LHC with hydrodynamical models suggest the matter formed in these collisions behave as nearly perfect fluid. Hydro calculations indeed indicate that,  in spite of the enormous energy and pressure fleetingly produced in these systems, they are characterized by small shear and bulk velocities per unit of entropy, commonly denoted  $\eta/s$ and $\zeta/s$, respectively. With the emergence of new measurement techniques and substantial theoretical advances, much progress has been recently achieved in  measuring $\eta/s$ and $\zeta/s$ with increasing precision \cite{flow1,flow2,flow3,flow4,flow5,flow6,flow7,flow8}. Shear and bulk viscosities are however only two of the many properties that characterize the bulk properties of systems and the matter they consist of. Other properties, such as the specific heat, the isothermal  compressibility, and various chemical potentials are also interest to fully characterize the properties of QGP matter and the nature of the phase transition from QGP to a hadron gas. There is hope, however, that such properties can be measured based on 
event-by-event fluctuations of ``global event observable`` such the particle multiplicity detected in a specific kinematic range, the mean transverse momenta ($\langle p_{\rm T} \rangle$), the total transverse  energies ($E_{T}$), as well as fluctuations of conserved quantities such as the net-charge and net-baryon
number of particles produced in collisions. Nominally, event-by-event fluctuations of net-charge and net-baryon number are proportional to the QGP susceptibilities and measurements of these fluctuations shall provide considerable insight onto the properties of the QGP and the phase transition. However, technical difficulties arise in the evaluation of these fluctuations because net charge and net baryon fluctuations are constrained by charge and baryon conservation and are  additionally strongly influenced by production and transport mechanisms. Similarly, matter properties such as the isothermal compressibility and the heat capacity, which  are nominally related to fluctuations of the produced particle multiplicity and the average (event-wise) particle momentum, are also challenging to extract because energy-momentum conservation, charge conservation, etc, also constrain fluctuations of the produced particle multiplicity and the momentum average of these particles. There is thus a need to learn how to account and remove known ``trivial" effects 
in measurements of such fluctuations. To this end, it is useful to consider predictions of  pQCD inspired elementary  collision models such as Pythia8 and Herwig\cite{pythia8,herwig}.  


The existence of a critical point at the QCD phase transition has been predicted to be 
associated with large event-by-event fluctuations in the aforementioned global observables \cite{fluct1,fluct2}. The relative fluctuations of the observables  in A$-$A collisions have been found to be smaller  compared to those in p$-$p collisions. The origin of fluctuations in both the systems is quite different and their measurement can provide information about the event-by-event 
initial state and final state fluctuations. The presence of thermal equilibration in A$-$A collisions makes it tricky to extract the information about the initial state from final state fluctuations. 
Therefore,  the basic goal in heavy ion collisions have been to relate the event-by-event fluctuations of the final state to the thermodynamic properties of the medium. 

The experimental measurement of event-by-event fluctuations in charged particle multiplicities in recent past searched for direct evidence of the QCD phase transition in different collision systems at different
energies \cite{wa98,na49,phenix1,phenix2}. If one considers the medium formed in relativistic nuclear collisions as a Grand Canonical Ensemble,  the isothermal compressibility ( $k_{T}$) of the system can be  related to the variance of the particle multiplicity near mid-rapidity as \\
\begin{equation}
  \langle (N  - \langle N \rangle)^{2} \rangle = \frac{k_{B} T \langle N \rangle^{2}}{V}k_{T} 
\end{equation}
 where  $\langle N \rangle$ is the mean multiplicity,  $k_{B}$ is the Boltzmann constant , T is the temperature and V is the volume of the system .
Thus, measurements of charged particle multiplicity fluctuations are expected to be a sensitive probe for critical behavior. 
A recent measurement by ALICE experiment on centrality dependence of  multiplicity fluctuations yielded a value of $k_{T} = 27.9$ $fm^{3}/GeV$ where the background fluctuation was estimated from the participant model \cite{alice, isothermal}.
In this work,  the multiplicity fluctuations of charged particles are obtained in Au$-$Au collisions at $\sqrt{s_{NN}}$ = 200 GeV and Pb$-$Pb collisions at $\sqrt{s_{NN}}$ = 2.76 TeV and 5.02 TeV using the Angantyr model of Pythia 8. The fluctuations are estimated over a large range of centralities and the effect of acceptance is presented by estimating  the observable in three different pseudo-rapidity intervals. The obtained results are also compared with the expectations of simple participant model and are expected to provide a baseline for current and future measurements and any significant deviation in 
experimental data might suggest the contribution of collectivity and final state effects.  

\section{Multiplicity Fluctuation Observable}

The relative fluctuation, $\omega_{X }$ of an observable X is defined  as\\
\begin{equation}
\omega_{X} =  \frac{\langle X^{2} \rangle -  \langle X \rangle^{2}}{ \langle X \rangle}
\end{equation}
where $ \langle . \rangle$ is the expectation value of the observable in the given event ensemble.
Fluctuations in the number of charged particles, $N_{ch}$, can be quantified as  the scaled variance of the charged particle multiplicity distribution \\
\begin{equation}
\omega_{ch} =  \frac{\langle N_{ch}^{2} \rangle - \langle N_{ch} \rangle^{2}}{ \langle N_{ch} \rangle}
\end{equation}

The measured value of $\omega_{ch}$ has contributions  originating from  both  statistical  as well as dynamical effects. In order to extract the  information from dynamical fluctuations and associated physical mechanism, one needs to clearly understand the different sources of fluctuations such as impact parameter fluctuations,  effect of limited acceptance of detector, fluctuations in the number of primary collisions, effect of resonance decays , effect of re-scattering of secondaries and other sources of correlations.

In the present work, the estimation of  fluctuations is also carried out using a simple participant superposition model. This model assumes that the nucleus-nucleus collision  as the sum of contributions 
from many sources created in the early stage of the interaction. The multiplicity fluctuations has contributions largely due to different impact parameters. The density fluctuations within the nucleus will make this contribution non-zero even if we consider a narrow  impact parameter window. The quantum fluctuations in the nucleon-nucleon cross sections can also lead to fluctuations in the number of particles produced by each source. These contributions are related to the initial volume of the interacting system and can lead to fluctuations in the number of participating nucleons, $N_{part}$ and can be related to the initial size of the system.
Therefore,  the fluctuations in $N_{ch}$ will have contributions due to fluctuations in $N_{part}$  as well as the fluctuations in the number of particles produced per participant ($n$). Hence, one can express the multiplicity fluctuations in the absence of correlations between the multiplicities produced by different 
sources by the following expression \\
\begin{equation}
\omega_{ch, parti} =   \omega_{n}  + \langle n \rangle \omega_{N_{part}}
\end{equation}

Although the particle production in central collisions is affected by nucleon-nucleon scattering, re-scatterings between produced particles, 
and other effects, the  estimations from such calculations and comparison with experimental data might divulge the validity of the superposition of nucleon-nucleon interactions in the case of heavy-ion collisions. One should also consider that $\omega_{n}$ will have a strong dependence on the acceptance of the detector.  

{\bf Estimation of  $\omega_{N_{part}}$ :}\\
The fluctuations in the impact parameter can reflect in the fluctuations in the number of participants. It has been estimated using the Pythia8 Angantyr model. 
The quantity $\langle n \rangle$ can be calculated  by taking the ratio of the mean charged particle multiplicity for a given acceptance to the
mean number of participants for the same centrality range.

{\bf Estimation of  $\omega_{n}$ :}\\
This term gives the fluctuation in the number of particles produced per participant. One needs to estimate the fraction $f$ which is the ratio of the mean number of particles per participant accepted within the acceptance under study  ($ \langle n \rangle$) to the mean number of total particles produced per participant ($\langle m \rangle$). While each participant produces $m$ charged particles, only a smaller fraction $f = \langle n \rangle/ \langle m \rangle$ are 
accepted.  Without carrying out a detailed analysis of the acceptance, one can make a simple statistical estimate assuming that the particles are accepted randomly, in which case $n$ is binomially distributed with mean $mf$ and $\sigma^{2}_{n} = m f (1  - f )$ for  a fixed $m$. Therefore, the fluctuations in the $n$ accepted particles is given as \\
\begin{equation}
\omega_{n} =   1 - f  + f\omega_{m}  
\end{equation}

\section{Analysis Method}
In the present study, the default heavy-ion model of Pythia 8, Angantyr model  has been used to generate 
Pb$-$Pb events at $\sqrt{s_{NN}}$ = 2.76 TeV and 5.02 TeV and  Au $-$ Au collisions at $\sqrt{s_{NN}}$ = 200 GeV.
The model is an extension of p$-$p collisional dynamics to nucleon-nucleus (p$-$A) and
nucleus-nucleus (A$-$A) collisions in the Pythia 8 event generator \cite{angantyr}.
For each A$-$A collision, the nucleons are  randomly distributed as per Wood-Saxon distribution in 
the impact parameter space and the formulation of individual  nuclear interactions is 
based on the Glauber model. The number of wounded nucleons are
estimated from Glauber formalism with Gribov corrections to the diffractive excitation of the 
individual nucleon. The Glauber calculations include the fluctuations of nucleons in the projectile and target nuclei.  
The model includes two  interaction methodologies for the projectile and 
target nucleons. In the first one, the interactions are treated as 
non-diffractive (ND) p$-$p collisions and are called primary ND interactions  while in the second case, 
a wounded projectile nucleon is allowed to have ND interactions with other target nucleons. These type of interactions are called  
secondary ND collisions. Depending on their interaction probability, various interactions between wounded nucleons in projectile and target are referred as elastic, ND, secondary ND, 
single-diffractive and double-diffractive interactions. The full Pythia8 machinery is used for parton-level event generation 
for these primary ND interactions (as well as diffractive interactions). 
Later, all the sub-events are stacked together to build an exclusive final state of the considered 
heavy-ion collision. The sub-events are treated independently where
hadrons are produced using the Lund string fragmentation model. As the model is devoid of collective effects and  
does not assume the  formation of any thermalised medium, it can act as a 
baseline model to understand the observables and the associated background which are affected by medium effects.

In the present study,  the transverse momentum, $p_{T}$  range of the charged particles is selected from $0.2 < pT < 2$ GeV/c. The event-by-event charged particle multiplicity 
has been determined for $|\eta| < $  0.5,  0.8  and 1.5 to study the dependence of the multiplicity fluctuations on acceptance.
The events are classified into different centrality classes using the two different centrality estimators, namely,  the total charged particle multiplicity obtained in the $\eta$ 
range $1.6 \leq |\eta| \leq 2.5$  ( CENT$\_$METH1) and the sum of transverse energy of all neutral particles produced (CENT$\_$METH2). The study is 
important as the variance of the charged particle multiplicity distribution is influenced by the centrality estimating method and may play an important role in observed 
scaled variance of charged particle multiplicity distributions. The centrality determination methods are chosen such that the observables are independent of methods and thus there is no bias due to selection of particles in the same kinematic acceptance.

\begin{figure*}
\begin{center}
\includegraphics[scale=0.78]{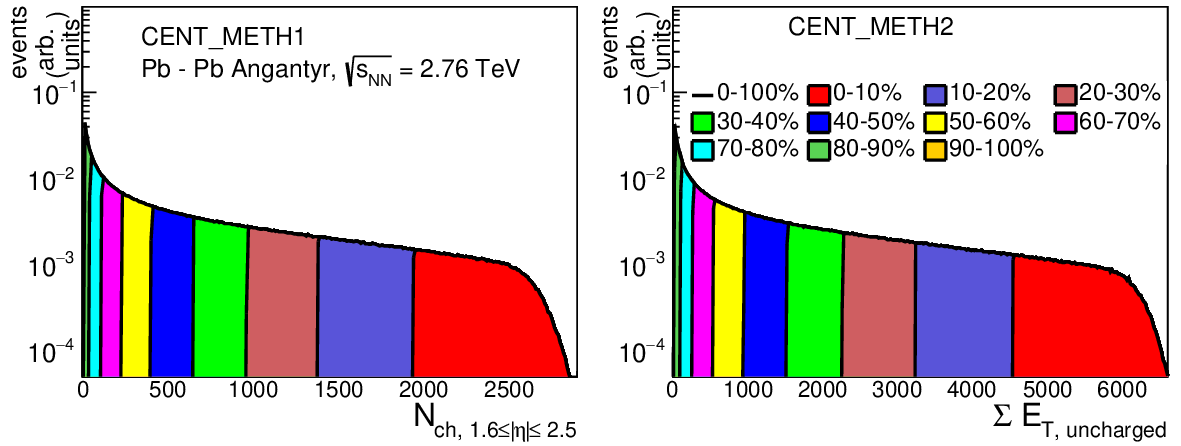}
\includegraphics[scale=0.78]{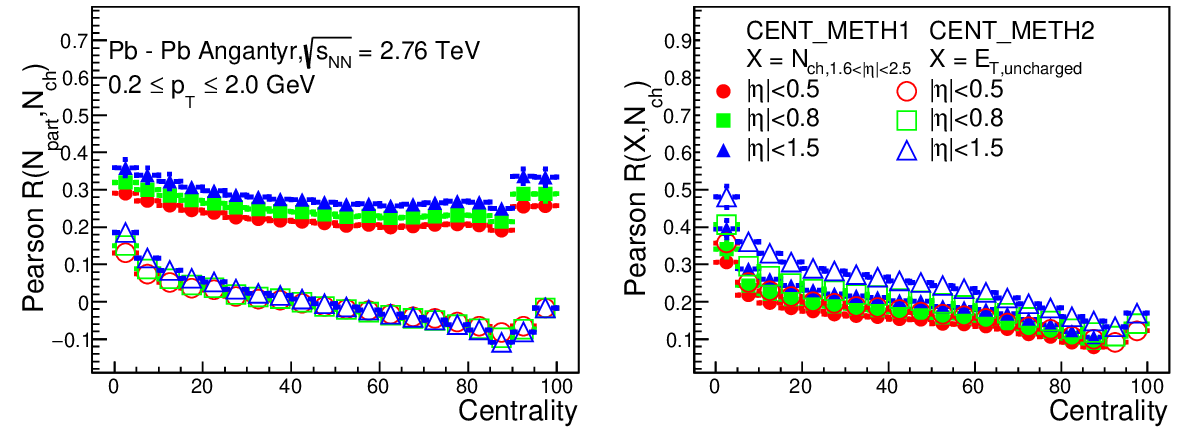}

\caption{(Color online)  [Upper panels] : The distribution of charged particle multiplicity in the $\eta$ 
range $1.6 \leq |\eta| \leq 2.5$ (left panel) and the distribution of transverse energy of neutral particles (right panel) in Pb$-$Pb collisions at  $\sqrt{s_{NN}}$ = 2.76 TeV. The centrality division has also been shown with different  coloured regions.
[Bottom panels]: The Karl Pearson correlation coefficient for number of participant nucleons ($N_part$) and Charged particle multiplicity for different centrality classes (left panel). The Karl Pearson coefficient for charged particle multiplicity and centrality determination estimator (X)  for three different acceptance regions }
\label{fig1}
\end{center}
\end{figure*}

The distribution of the  considered observables for centrality distribution and their division into different centralities is shown in Figure \ref{fig1} for Pb$-$Pb collisions at $\sqrt{s_{NN}}$ = 2.76 TeV.
An additional check was performed to check the correlations between the  obtained charged particle multiplicity in different $\eta$ regions  and the number of participants 
obtained by the two different methods by calculating the Pearson Coefficient between the variables. The correlation coefficient was also determined between the  number of 
charged particles in the analysis range and the centrality estimators ( charged particle multiplicity ($1.6 \leq |\eta| \leq 2.5$ ) and transverse energy of neutral particles)  for different centrality classes. This is shown in lower panels of Figure \ref{fig1}  for Pb$-$Pb collisions at $\sqrt{s_{NN}}$  = 2.76 TeV.  The lower left panel shows a considerable difference between the two methods and it is more pronounced for peripheral collisions while there is a good agreement in the right panel. One can observe that the correlation coefficient between the shown observables is around 0.3 for central collisions and it decreases from central to peripheral collisions.

A given centrality class is a collection of events that comprises of a range of impact parameters or number of participating nucleons, $N_{part}$. This results in additional fluctuations in the number of 
produced particles within each centrality class. In order to infer about the genuine fluctuations arising from different dynamical processes, one needs to ensure that the fluctuations in 
$N_{part}$ within a centrality class is minimal.The fluctuations arising due to wider centrality bin width is corrected or minimised by dividing the centrality bin into smaller bins and weighing the 
corresponding moments as \\
\begin{equation}
X = \frac{\sum \limits_{i} n_{i}X_{i}}{\sum \limits_{i} n_{i}}
\end{equation}
where the index $i$ runs over each multiplicity bin, $X_{i}$ represents various measured observable for the
$i^{th}$ bin, and $n_{i}$ is the number of events in the $i^{th}$ centrality bin. This method is commonly known as centrality bin width correction method and the final results are presented for a centrality bin width of 5\% after implementing  the centrality bin-width corrections of 1\% bin width \cite{binwidth,binwidth1}.

\section{Results and Discussion}
\begin{figure*}

\begin{center}
\includegraphics[scale=0.5]{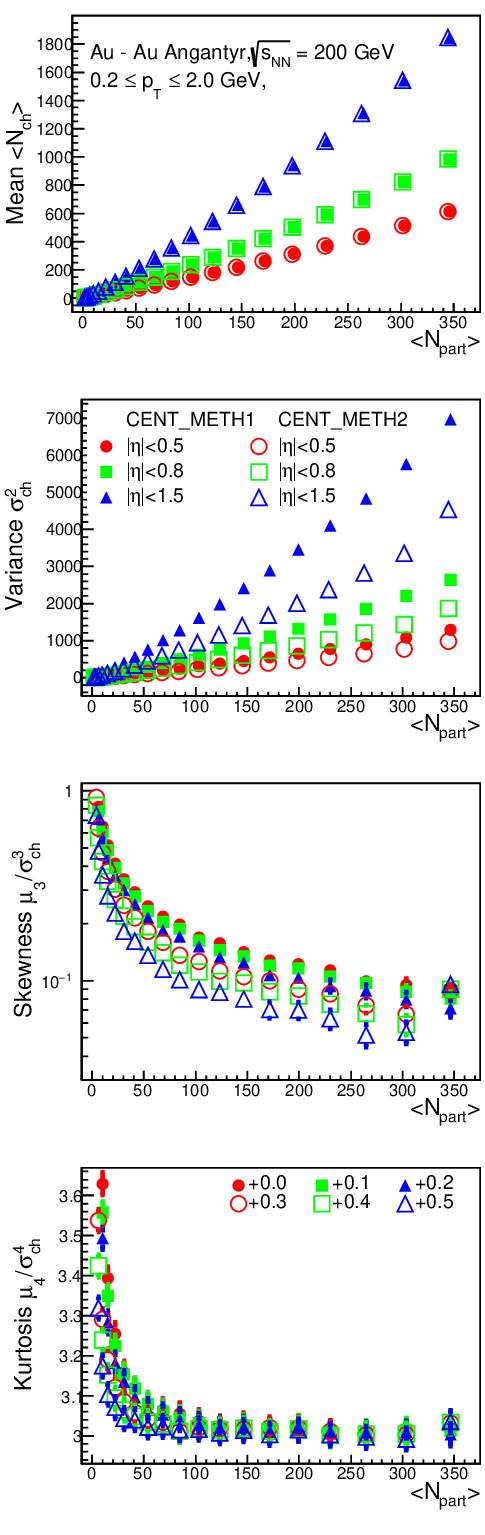}
\includegraphics[scale=0.5]{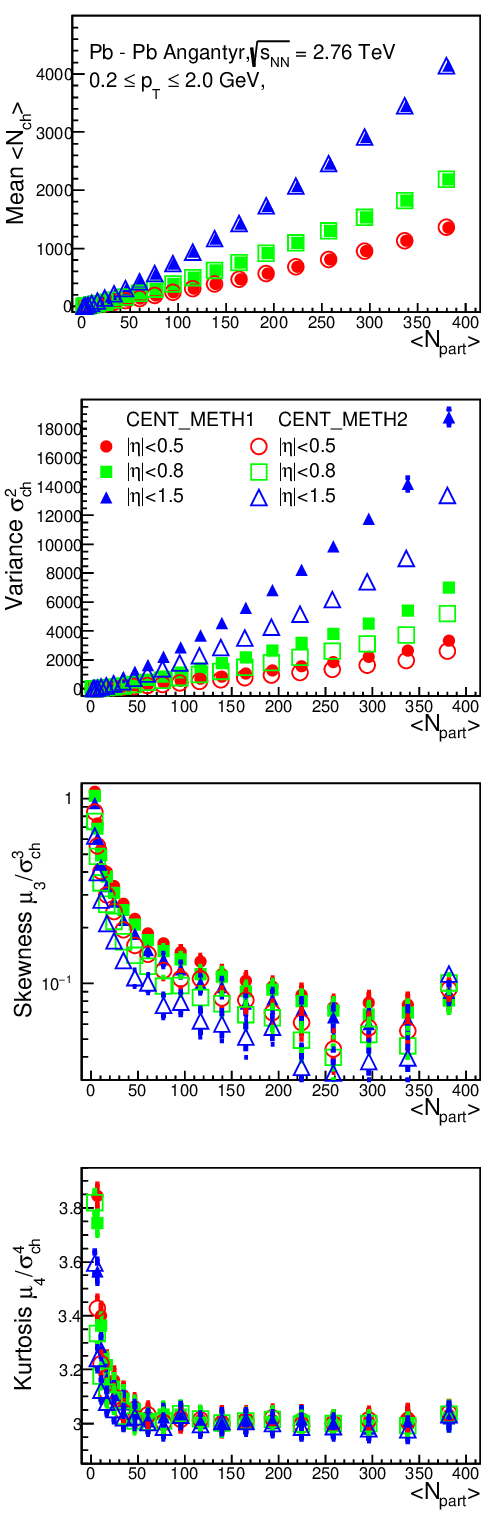}
\includegraphics[scale=0.5]{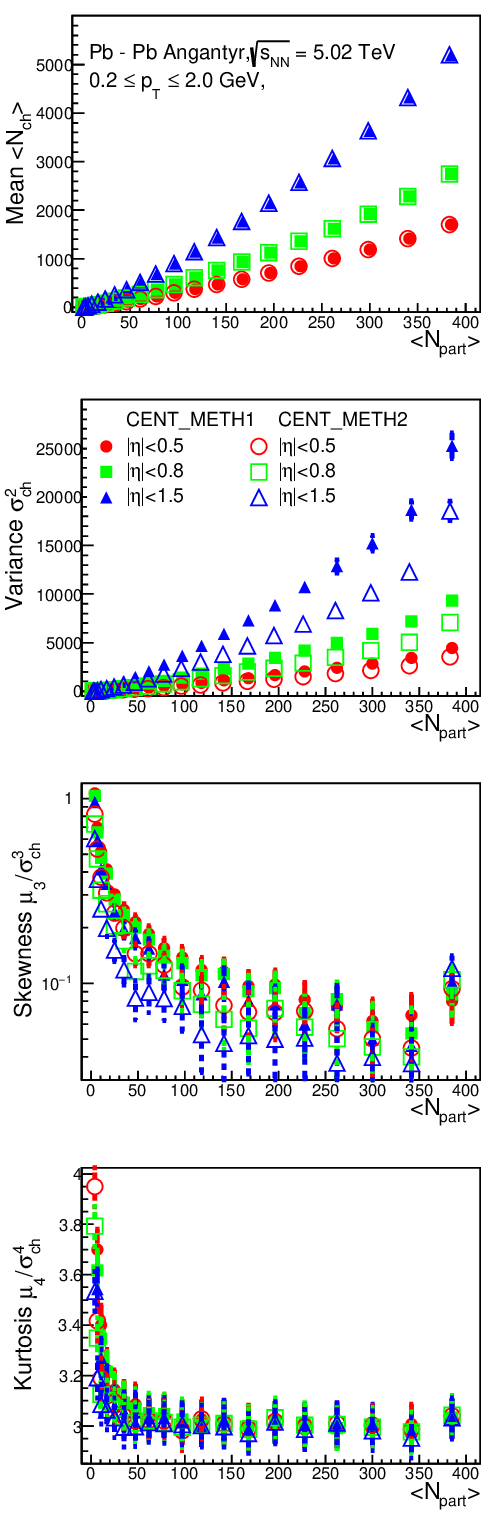}
\caption{(Color online) The variation of  $\langle N_{ch} \rangle$, variance, skewness and kurtosis of charged particle multiplicity with $ \langle N_{part} \rangle$( centrality) for different $\eta$ ranges estimated for two centrality determination method. The solid and open markers are for two different methods of centrality estimation.}
\label{fig2}
\end{center}
\end{figure*}

Figure  \ref{fig2} compares  the  mean, variance, skewness and kurtosis  of the charged particle multiplicity distributions obtained for different $\eta$ ranges as a function of 
centrality using two different methods of centrality determination.  It can be observed that the variation of $\langle N_{ch} \rangle$ with $\langle N_{part} \rangle$ shows an expected increasing trend 
with $\langle N_{part} \rangle$ and there is a good agreement for the values obtained through the two centrality determination methods. 
However, the same is not true for the trend of variances of charged particle multiplicity distribution and the disagreement increases with an increase in $\eta$ interval. 
The observed difference in principle can affect the values of  $\omega_{ch}$. The  skewness and kurtosis decreases with  $\langle N_{part} \rangle$ and is in agreement with the 
expectations of central limit theorem. 

\begin{figure}
\begin{center}
\includegraphics[scale=0.35]{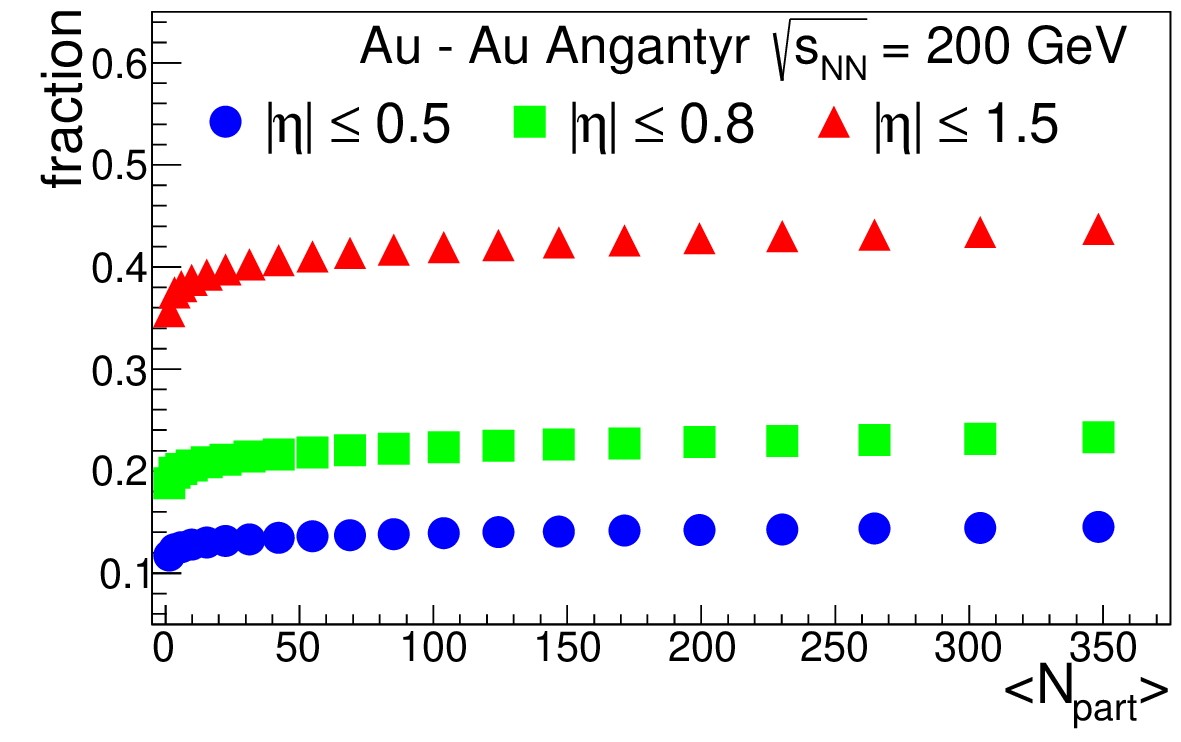}
\includegraphics[scale=0.35]{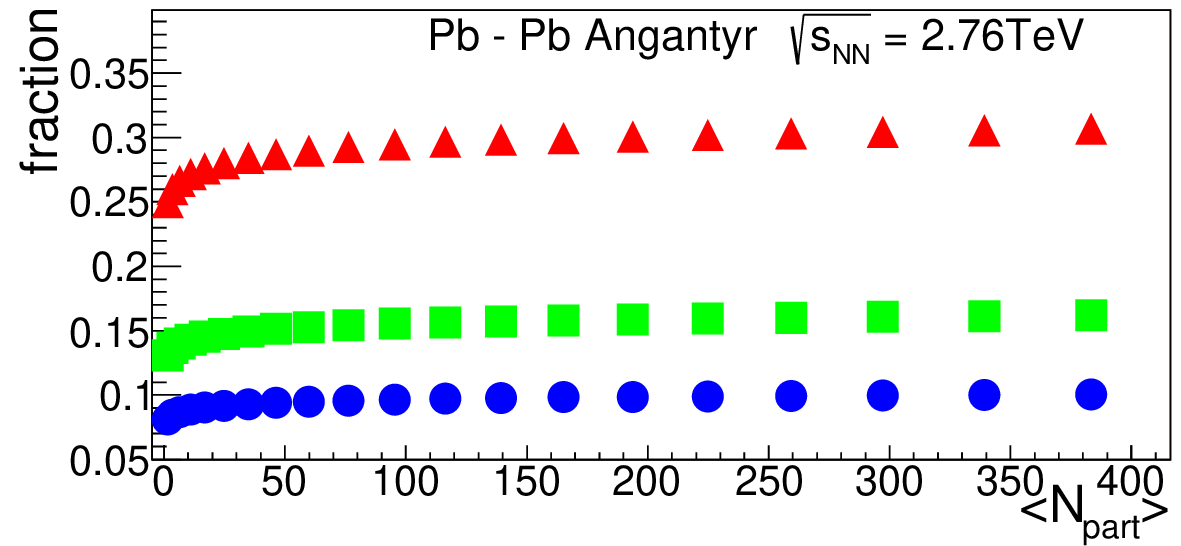}
\includegraphics[scale=0.35]{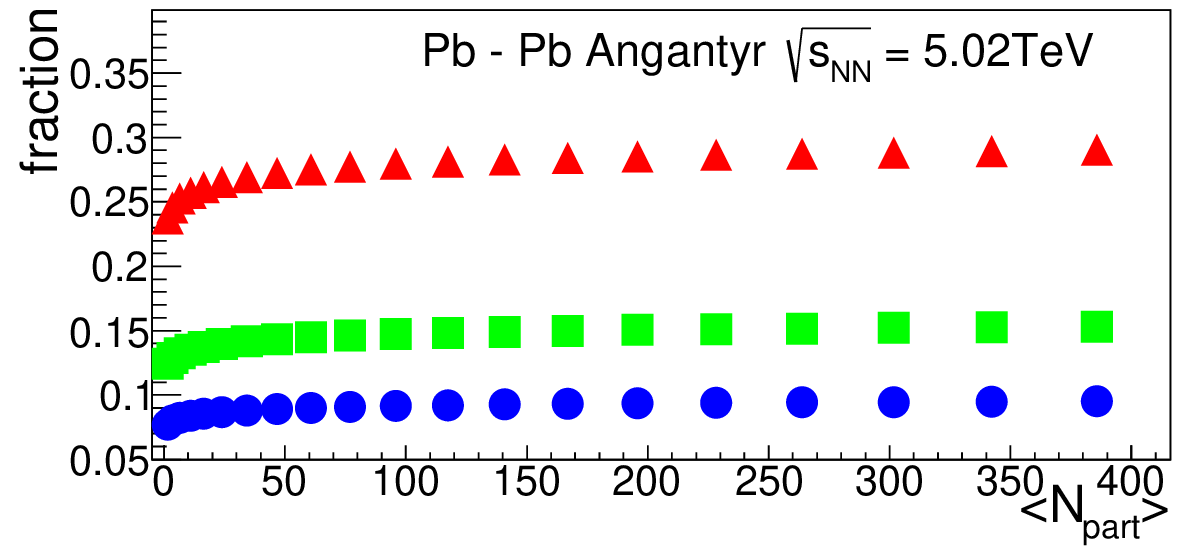}
\caption{(Color online) The variation of  $f$  with $\langle N_{part} \rangle$( centrality) for different $\eta$ ranges estimated for the three collision systems.}
\label{fig3}
\end{center}
\end{figure}

In order to estimate the $\omega_{ch}$ values from the wounded participant model, one needs to obtain the values of $f$, fraction of charged particles accepted to the total number of charged particles produced. The obtained value is used in Equation [5]  to obtain the values of $\omega_{n}$. The variation of this fraction with centrality for different pseudo-rapidity intervals  is shown in Figure \ref{fig3}. It can be observed that $f$ shows a slightly increasing trend with centrality. However, the values decrease with beam energy for similar pseudo-rapidity acceptance.

\begin{figure*}
\begin{center}
\includegraphics[scale=0.8]{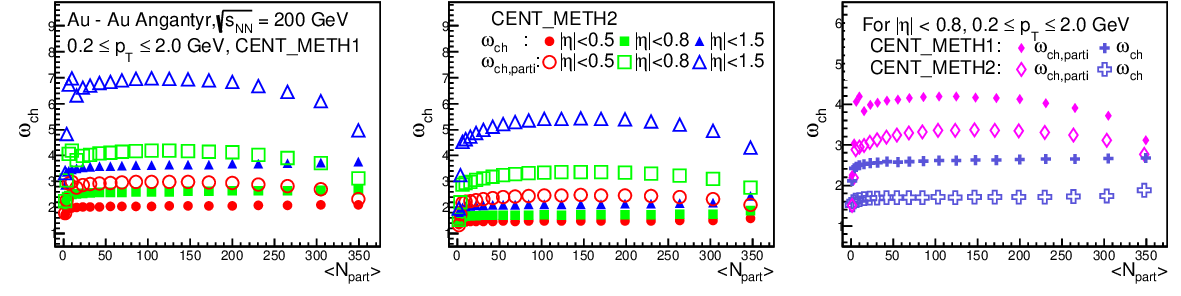}
\includegraphics[scale=0.8]{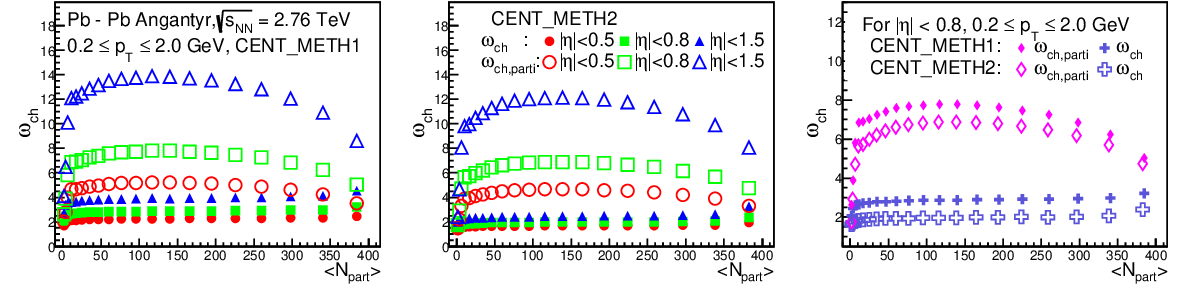}
\includegraphics[scale=0.8]{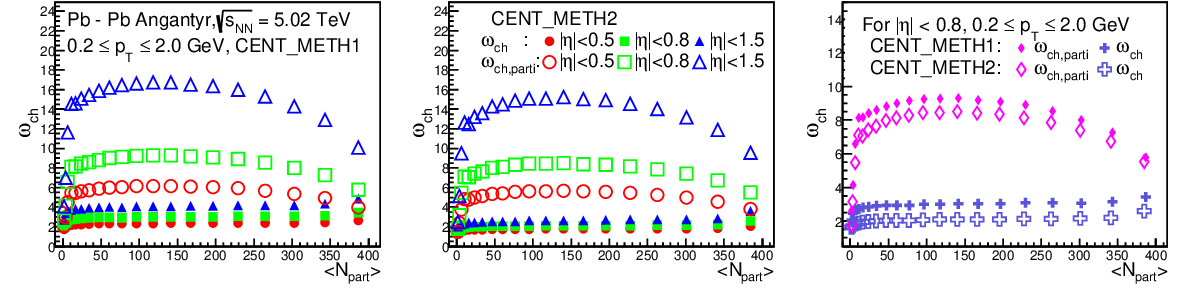}
\caption{(Color online) The variation of  $\omega_{ch}$ with centrality for different collision systems at different centre of mass energies. The  $\omega_{ch}$ values have been compared to  the 
expectations of a simple model of participant overlap, $\omega_{ch,parti}$.}
\label{fig4}
\end{center}
\end{figure*}

The variation of $\omega_{ch}$  obtained from the Angantyr model with centrality is compared with the expectations from  the wounded participant model in Figure \ref{fig4}. The comparison is also shown for different $\eta$ ranges.
The left hand side panels show that the values obtained from the Angantyr model are consistently lower than the geometrical fluctuations. There is no significant centrality and beam energy dependence. However, one can observe that the central values show an increasing trend with centrality for $\omega_{ch}$ while it is opposite for the ones obtained from the simple participant model. The values in wider acceptance intervals are higher for all the energies and systems studied.  The middle panel shows the same for a different centrality estimator and the trend remains the same. The right panel compares the estimations 
from two different centrality estimator  methods and one can observe a noticeable difference between the obtained values. This observation can be  attributed to different 
values of correlation coefficient between the charged particle multiplicity and the number of participants obtained from the two methods of 
centrality estimation as shown in Figure \ref{fig1}.  This indicates towards the sensitivity of the multiplicity fluctuations studies on  methods of centrality estimation.   
The values of the scaled variance obtained from the model remain significantly above the random (Poisson) expectation of 1.0.

\begin{figure}
\begin{center}
\includegraphics[scale=0.35]{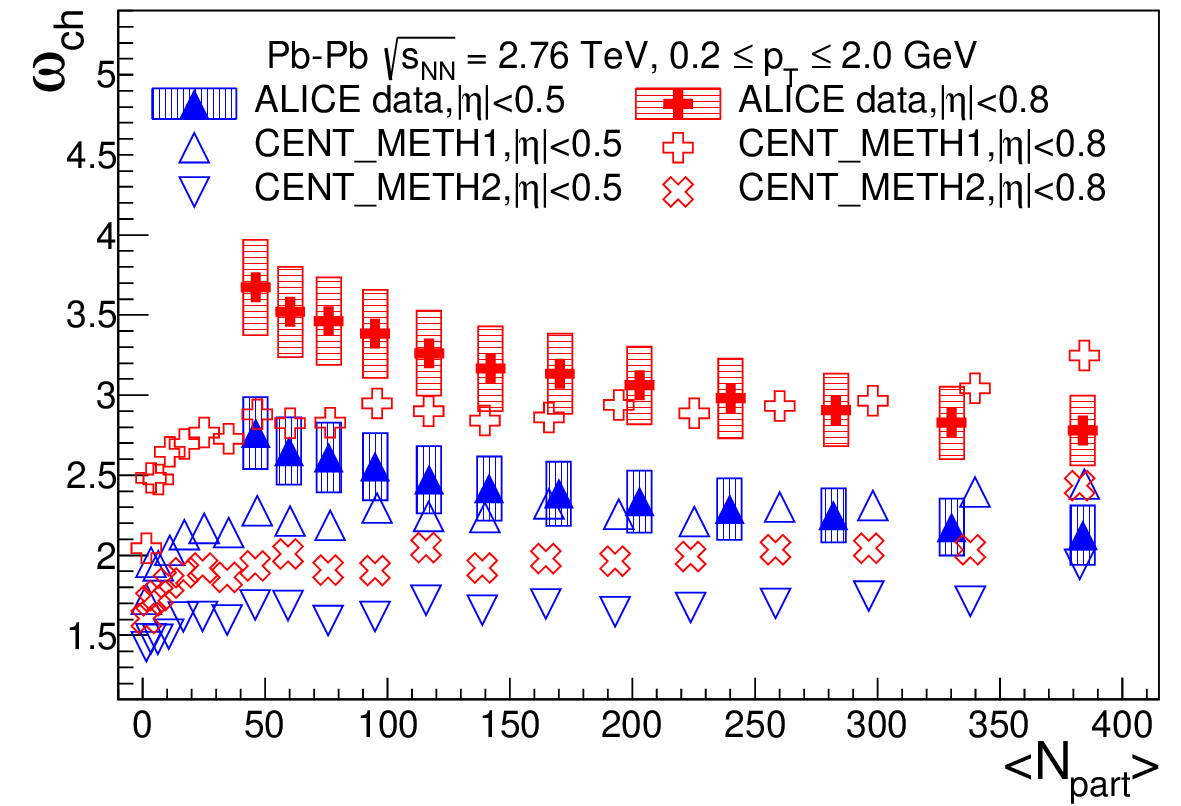}
\caption{(Color online) The variation of  $\omega_{ch}$ as measured by ALICE experiment is compared with the predictions from Angantyr model for two different $\eta$ regions.}
\label{fig5}
\end{center}
\end{figure}
A comparison of the $\omega_{ch}$ values obtained from Angantyr model  and that measured by ALICE experiment \cite{alice} has been shown in Figure \ref{fig5} for Pb$-$Pb collisions at $\sqrt{s_{NN}}$ = 2.76 TeV. One can observe that the model under-predicts the data in lower centralities while the central values are in reasonable agreement for high values of  $\langle N_{part} \rangle$. However, the trend shown 
for evolution with centrality is quite opposite. The model predicts a slightly increasing trend with $\langle N_{part} \rangle$ while the data shows a decreasing trend.

\begin{figure}
\begin{center}
\includegraphics[scale=0.3]{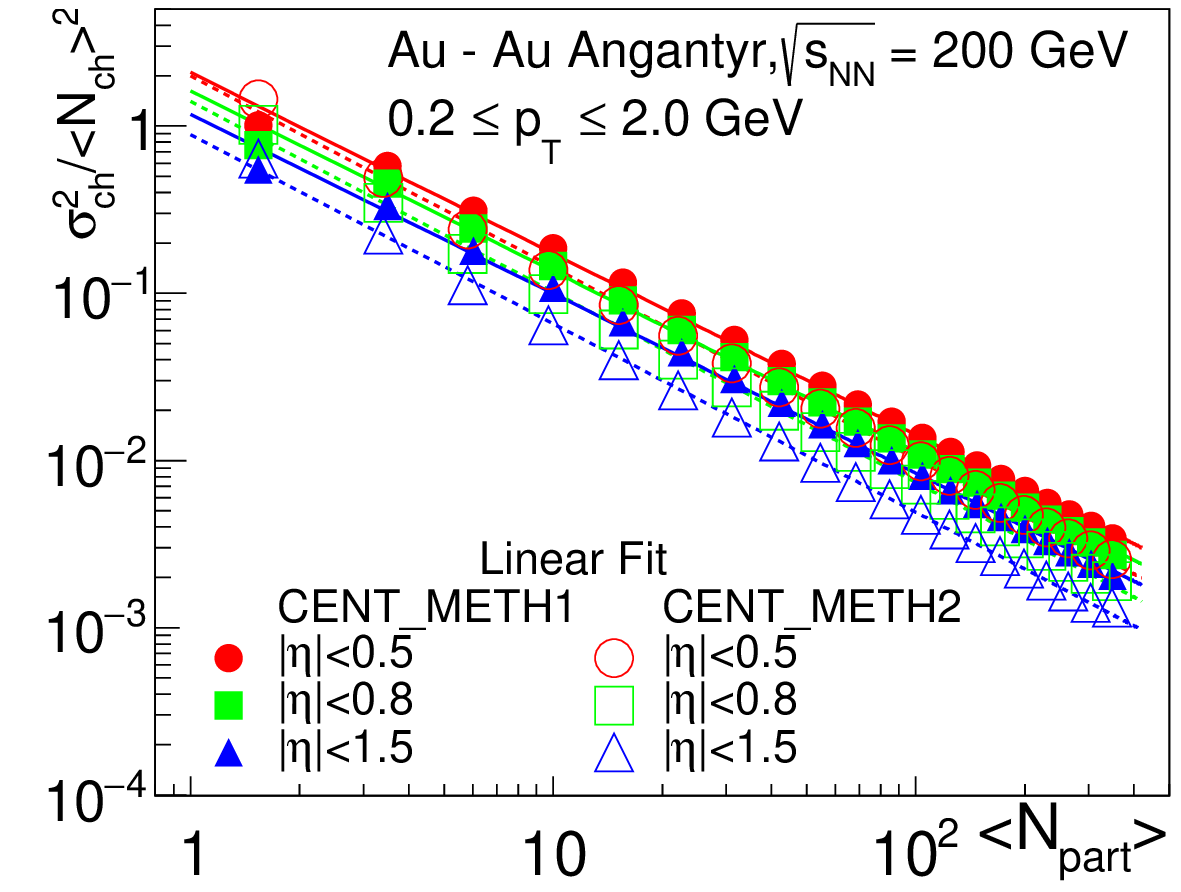}
\includegraphics[scale=0.3]{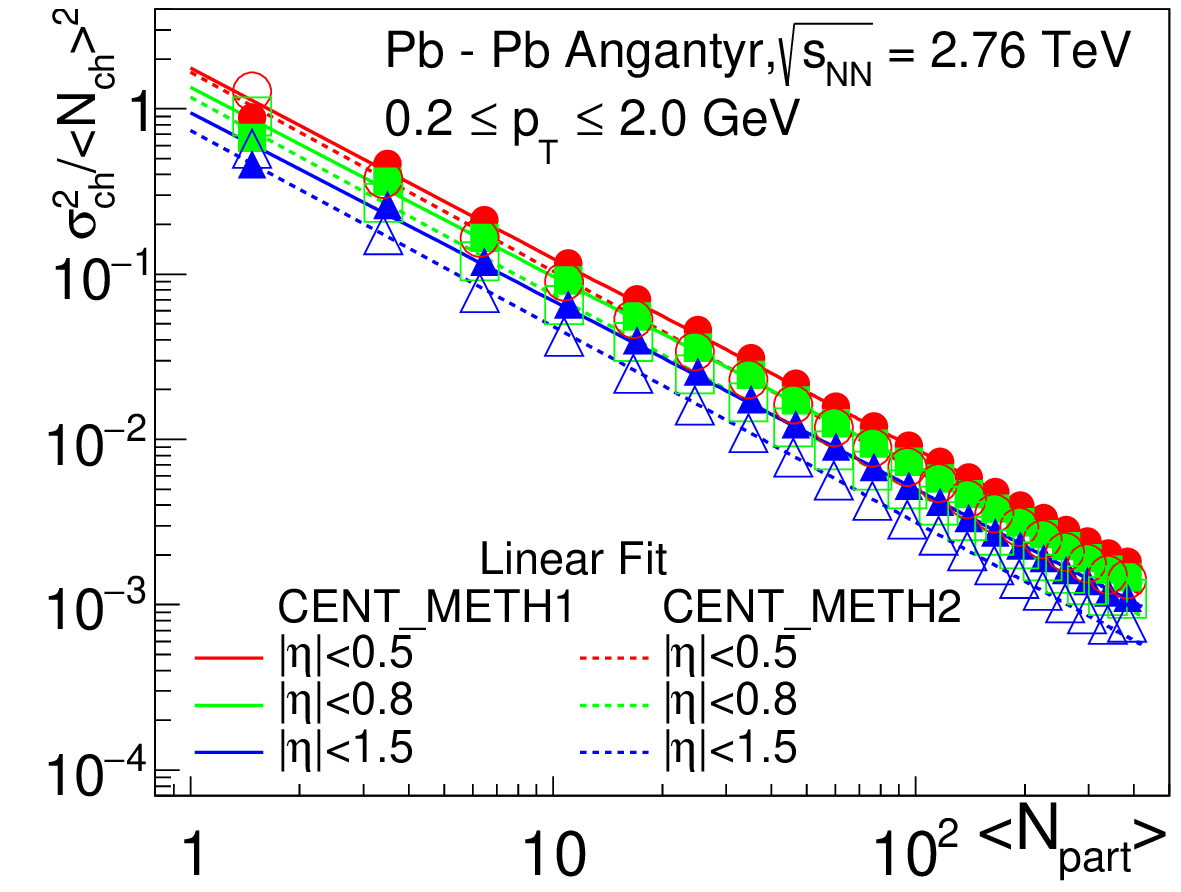}
\includegraphics[scale=0.3]{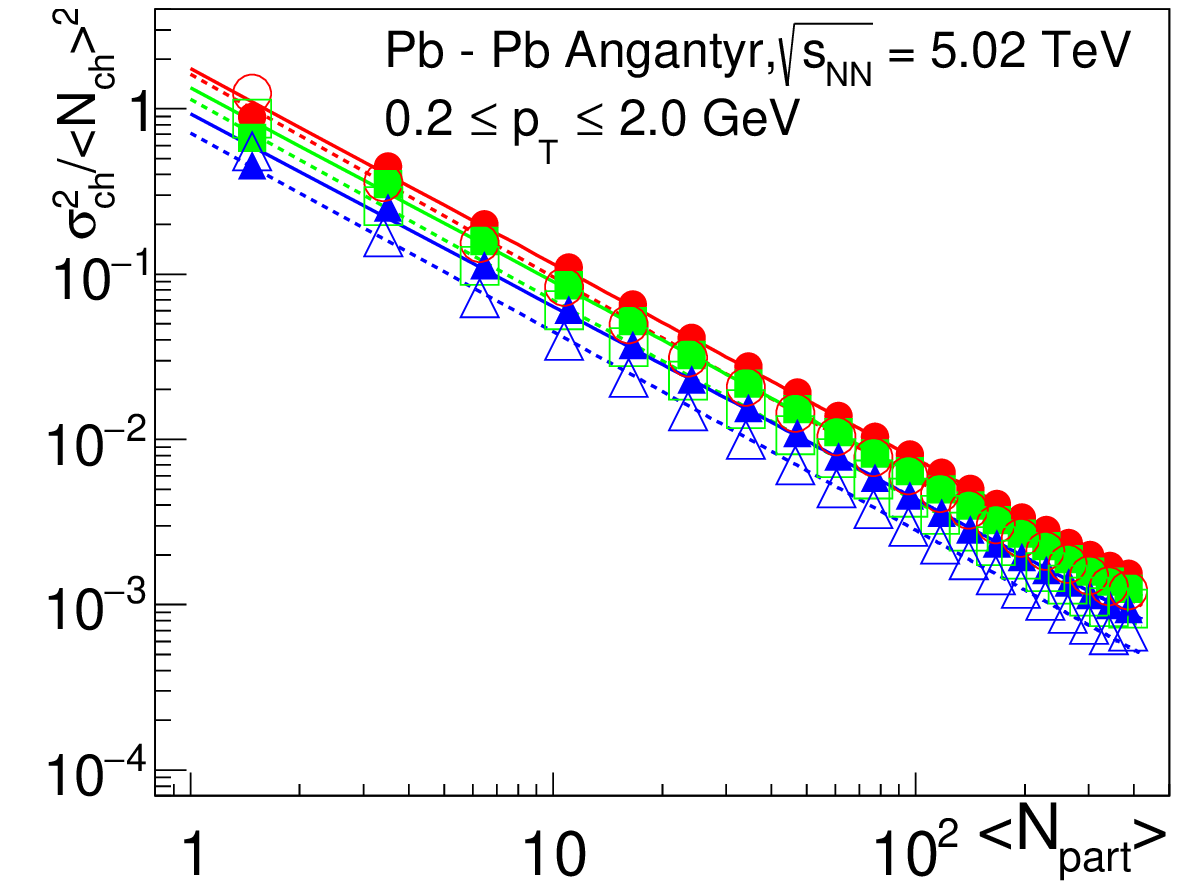}
\caption{(Color online) The $\sigma^{2}/\mu^{2}$  for charged hadrons in the transverse momentum range $0.2< p_{T}<2.0$ GeV/c for different collision systems as a function of $<N_{part}>$. The solid line shows the fit function $ln R  = a_{1} ln(N_{part}) + a_{0}$ to the obtained values of R.}
\label{fig6}
\end{center}
\end{figure}

Figure \ref{fig6}  shows the fluctuations in terms of  $R =   \sigma^{2}/\mu^{2}$ as a function of the $\langle N_{part} \rangle$ for all the three systems. 
The dependence of the observable can be best described by the power law scaling as a function of $\langle N_{part} \rangle$. The observed scaling is independent of the system and beam energy
and is in agreement with the results observed from different experiments.
The details of the extracted fit parameters  ( $ln R  = a_{1} ln( \langle N_{part} \rangle) + a_{0}$ ) are shown in Table \ref{table1}. 

The present study provides an estimation of the $\omega_{ch}$ obtained from Angantyr model and it will be a relevant baseline to the measured values from LHC data as the model does not consider 
the formation of a deconfined medium.  An estimate from the participant model indicates that one needs to consider additional sources to account for background estimation of multiplicity fluctuations.
\begin{table}[h!]
\centering
 \begin{tabular}{c|c|c|c|c|c}
 \hline
 $\sqrt{s_{NN}}$ &  $\eta$ range &  $a_{0}$  & $a_{1}$   & $b_{0}$  & $b_{1}$ \\ [0.5ex]
 \hline\hline
  Au+Au  & $|\eta| < 0.5$ & 0.74   & -1.084 & 0.69 & -1.14  \\
200 GeV & $|\eta| < 0.8$ & 0.48   & -1.078  & 0.34 & -1.14\\
                               & $|\eta| < 1.5$ & 0.16   & -1.073  & -0.12 & -1.13 \\	
\hline                                                     
  Pb+Pb & $|\eta| < 0.5$ &0.56   & -1.15  & 0.51 & -1.20 \\
  2.76 TeV                             & $|\eta| < 0.8$ &0.29   & -1.14 & 0.16& -1.20  \\
                               & $|\eta| < 1.5$ &-0.05   & -1.14  & -0.30 & -1.19 \\	

\hline  

  Pb+Pb  & $|\eta| < 0.5$ &0.56   & -1.18   & 0.48 & -1.23\\
   5.02 TeV                           & $|\eta| < 0.8$ &0.29   & -1.17 & 0.13. & -1.22  \\
                               & $|\eta| < 1.5$ &-0.07   & -1.16  & -0.34 & -1.20 \\	
\hline\hline
\end{tabular}
\caption{A list of fit parameters for  different collision systems. $a_{0}$ ($b_{0}$) and $a_{1}$ ($b_{1}$) are the intercept and the slope parameters extracted  from the  
fit in CENT$\_$METH1(CENT$\_$METH2)}
\label{table1}
\end{table}

\section{Summary}

A detailed study of event-by-event fluctuations in the multiplicities
of charged particles has been carried out using the Pythia8 Angantyr model 
for Au$-$Au collisions at $\sqrt{s_{NN}}$ =  200 GeV and Pb$-$Pb collisions at  $\sqrt{s_{NN}}$ = 2.76 TeV and 5.02 TeV.
The multiplicity fluctuation observable, $\omega_{ch}$ has been obtained for different centrality
intervals as well as for different pseudo-rapidity classes.  The centrality bin width correction has been implemented 
to take care of the contribution from impact parameter fluctuations.  A  small dependence on centrality method estimation is also observed 
which can be treated as a source of systematic error while estimating the $\omega_{ch}$ parameter. The observation was attributed to different 
values of correlation coefficient between the charged particle multiplicity and the number of participants obtained from the two methods of 
centrality estimation.  The fluctuations in multiplicities are not observed to vary significantly as a function of centrality and beam energy.  However, there is an 
observed centrality dependence of the multiplicity fluctuations of
charged particles  obtained from a simple participant model which shows a decrease with an increase in centrality.  A decrease in
$\eta$ acceptance results in decreased multiplicity fluctuations which can be understood by a simple statistical analysis. 
The observed variance scaled with the square of mean multiplicity obeys a power law scaling with the number of participants for 
the three collision systems and the scaling is in agreement with the measured experimental data. 

\section{Acknowledgements}
The authors would like to thank the Department of Science and Technology (DST), India for supporting the present work.


\begin{thebibliography}{100}
\medskip

\bibitem{flow1} D.~Teany,  Phys. Rev. {\bf C 68}, 034913, (2003).
\bibitem{flow2} R.~A.~Lacey  {\it et al.},  Phys. Rev. Lett.{\bf 98}, 092301, (2007).
\bibitem{flow3} H.-J. Drescher, A. Dumitru, C. Gombeaud, and J.-Y. Ollitrault, Phys. Rev. {\bf C 76}, 024905 (2007).
\bibitem{flow4} K. Dusling and D. Teaney, Phys. Rev. {\bf C 77}, 034905 (2008).
\bibitem{flow5} Z. Xu, C. Greiner, and H. St¨ocker, Phys. Rev. Lett. {\bf 101}, 082302 (2008).
\bibitem{flow6} D. Molnar and P. Huovinen, J. Phys. {\bf G 35}, 104125 (2008).
\bibitem{flow7} A. Adare  {\it et al.}, (PHENIX Collaboration),  Phys. Rev. Lett. {\bf 107}, 252301 (2011).
\bibitem{flow8} K. Aamodt  {\it et al.}, (ALICE Collaboration), Phys. Rev. Lett. {\bf 107}, 032301 (2011).
\bibitem{pythia8} T ~Sjostrand, S.`Ask, J.~R.~Christiansen, R. ~Corke, N.~Desai, P.~Ilten,S.~Mrenna, S.~Prestel, C.O.~Rasmussen , P.Z.Skands, Comput. Phys. Commun. {\bf 191},159-177 (2015).
\bibitem{herwig} J. ~Bellm, EPJ {\bf C} 76(4), 196 (2016).
\bibitem{fluct1} M. Stephanov, K. Rajagopal, and E. Shuryak, Phys. Rev. {\bf D 60}, 114028 (1999). 
\bibitem{fluct2} F. Karsch, S. Ejiri, K. Redlich, Nucl. Phys. {\bf A 774}, 619(2006)
\bibitem{wa98} M.M.~Aggarwal {\it et al.}, (WA98 Collaboration),  Phys. Rev. {\bf C 65}, 054912, (2002).
\bibitem{na49} C.~Alt {\it et al.}, (NA49 Collaboration),  Phys. Rev. {\bf C 75}, 064904, (2007). 
\bibitem{phenix1} S.S. ~Adler {\it et al.}, (PHENIX Collaboration), Phys. Rev.  {\bf C 76}, 034903 (2007). 
\bibitem{phenix2} A. ~Adare {\it et al.}, (PHENIX Collaboration), Phys. Rev.  {\bf C 78}, 044902(2008).
\bibitem{alice}	S. ~ Acharya {\it et al.}, (ALICE Collaboration), Eur. Phys. J {\bf C 81} 1032 (2021)
\bibitem{angantyr} C.  Bierlich,  G.  Gustafson,  L.  Lnnblad  and H. Shah, JHEP {\bf 1810}, 134 (2018).
\bibitem{isothermal} Maitreyee Mukherjee, Sumit Basu, Arghya Chatterjee, Sandeep Chatterjee, Souvik Priyam Adhya, Sanchari Thakur, Tapan K. Nayak, Phys. Letts.  {\bf B 784}, 1-5 (2018).
\bibitem{binwidth} T. Sugiura, T. Nonaka, and S. Esumi, Phys. Rev.  {\bf C 100}, 044904 (2019).
\bibitem{binwidth1} Maitreyee Mukherjee {\it et al.}, J. Phys. G: Nucl. Part. Phys. {\bf 43} , 085102 (2016).

\end{thebibliography}
\end{document}